\documentclass{aa}
\usepackage{graphicx}
\usepackage{txfonts}
\begin{document}


\title{The ESO Nearby Abell Cluster Survey
       \thanks{Based on observations collected at the European Southern
               Observatory (La Silla, Chile)}
}

\subtitle{IX. The morphology-radius and morphology-density relations \\ 
in rich galaxy clusters}
\author{T.~Thomas, P.~Katgert}
\institute{ Sterrewacht Leiden, The Netherlands}
\offprints{P.~Katgert}
\date{Received date; accepted date}
\markboth{The ESO Nearby Abell Cluster Survey IX}{The morphology-radius and 
morphology-density relations in rich clusters}

\abstract{
We study the morphology-radius and morphology-density relations for a
sample of about 850 galaxies (with $M_R \le -19.5$) in 23 clusters
from the ENACS (ESO Nearby Abell Cluster Survey). On the basis of
their radial distributions we must distinguish: (i) the brightest
ellipticals (with $M_R < -22$), (ii) the late spirals, and (iii) the
{\em ensemble} of the less bright ellipticals, the S0 galaxies and the
early spirals, which have indistinguishable distributions of projected
radial distance $R$. The brightest ellipticals are most centrally
concentrated, while the late spirals are almost absent from the
central regions; the radial distribution of the other galaxy classes
is intermediate. The previously found radial segregation of the
ellipticals thus appears to be due to {\em the brightest ellipticals
only}, while that of the spirals is due to {\em the late spirals
only}. \\ The morphology-density (MD-) relation was derived with two
measures of projected density: one using the 10 nearest neighbours
($\Sigma_{10}$) and another using only the nearest neighbour
($\Sigma_{1}$). In the $\Sigma_{10}$ MD-relation, only the classes of
early- and late-type galaxies show a significant difference, but the
different galaxy types within those classes are indistinguishable.
However, this result is affected by significant cross-talk from the
morphology-radius (or MR-) relation, as $\Sigma_{10}$ is strongly
correlated with $R$.  $\Sigma_{1}$ appears much less correlated with
$R$ and therefore the crosstalk from the MR-relation is much
smaller. As a result, the normal `ellipticals' (with $M_R \ge -22$),
the S0 galaxies and the early spirals {\em do have} different
$\Sigma_{1}$-distributions. {\em On average}, the 'normal' ellipticals
populate environments with higher projected density than do the S0
galaxies while the early spirals populate even less dense
environments. \\ We conclude that the segregation of the brightest
ellipticals and the late spirals is driven primarily by {\em global}
factors, while the segregation between 'normal' ellipticals, S0
galaxies and early spirals is driven mostly by {\em local} factors. We
discuss briefly the implications of these results in terms of
scenarios for formation and transformation of galaxies in clusters.
}

\maketitle

\section{Introduction}
\label{s-intro}

In the past thirty years many observers have studied the relation
between morphology and cluster environment. Oemler (1974), Melnick
\& Sargent (1977) and Dressler (1980) were the first to quantify 
differences in the projected distributions of galaxies of various
morphological types. Before this time it already was widely accepted
that the Hubble classification reflects a sequence of physical
properties. Yet, although the morphological classes appear to describe
fundamental properties of galaxies, it is not very clear how those are
determined by the (local or global) conditions in which a cluster
galaxy finds itself.

Luminosity segregation (i.e the fact that the projected distribution
of galaxies within a cluster depends on luminosity) was found by Rood
\& Tunrose (1968), Capelato et al.\ (1980) and Kashikawa et al.\
(1998). In addition, Beisbart \& Kesher (2000) found that bright
galaxies are more strongly clustered than faint galaxies and Biviano
et al. (2002, hereafter paper XI) found that luminosity segregation is
limited to the brightest ellipticals.

In addition, it was also found that there is a relation between
morphological type and projected density. This morphology-density
relation has been studied for local clusters (Dressler 1980, Goto et
al. 2003) as well as at intermediate redshifts (e.g. Dressler et
al. 1997, Fasano et al. 2000, Treu et al. 2003, and Nuijten et al.
2005). Detailed studies of morphological segregation in clusters at
low redshifts can provide a better understanding of the relations
between the morphological classes. Prugniel et al.\ (1999) showed that
galaxies likely to contain young sub-populations are preferentially
found in less dense environments, while Goto et al. (2003) found that
late disk galaxies avoid the dense central regions of clusters. At the
same time, the fraction of gas-poor galaxies increases and the
fraction of emission-line galaxies (ELG) decreases towards the dense
cluster center (Biviano et al.\ 1997, paper III; Solanes et al. 2001;
Dale et al.\ 2001; Thomas \& Katgert 2005, paper VIII).
 
The variation of the morphology-density relation with redshift adds
information on the evolutionary relationships between cluster galaxies
of different types and on possible transformation relations between
them. On the one hand, Goto et al. (2003) argue that the
morphology-density relations at $z = 0$ and $z = 0.5$ are very
similar. On the other hand, Treu et al. (2003), who made a detailed
study of the morphology-density relation in a cluster at $z = 0.4$,
and Nuijten et al. (2005), who studied the morphology-density relation
out to $z \sim 1$, find that the fraction of early-type galaxies in
the overdense regions increases towards lower redshifts.

The latter studies thus confirm the findings of Dressler et al.\
(1997) and Fasano et al.\ (2000) that the fraction of S0 galaxies in
clusters increases towards lower redshifts (but see Andreon 1998 and
Fabricant et al.\ 2000). Results from e.g.\ Poggianti et al.\ (1999)
and Jones et al.\ (2000) suggest that many early spirals have
transformed into S0 galaxies, possibly by impulsive encounters (Moore
et al.\ 1999). These results can be reconciled with the apparently
passive evolution of most early-type galaxies if the progenitor bias
is taken into account (van Dokkum \& Franx 2001).  Thus, early-type
galaxies that underwent star formation at $z\sim0.5$ (such as observed
by Ferreras \& Silk 2000) would not be identified as early-type
galaxies at that redshift.

In studies of the (evolutionary) relationships between cluster
galaxies of different types it is important to distinguish between
local and global processes. Sanroma \& Salvador-Sol\'e (1990), and
subsequently Whitmore et al.\ (1993) argued that the cluster-centric
radius, a global parameter, is the most fundamental parameter, because
they found a very strong correlation between morphology and
cluster-centric radius. However, Dressler et al.\ (1997) argued that
the morphology-density relation, which is probably the result of local
processes, is more fundamental since it is observed for both regular
and irregular clusters. 

One of the reasons for these different conclusions may be that it is
not trivial to separate global (radius) and local (density)
segregation, as density and radius are generally correlated. Dominguez
et al.\ (2001) tried to separate the two effects and concluded that in
the inner regions of clusters, segregation seems to depend mostly on
global parameters (cluster-centric radius or mass density), while in
the outer region of clusters segregation can be best described by
local parameters, such as projected galaxy density.

In this paper we use the galaxy types derived by Thomas \& Katgert
(2005, paper VIII) for galaxies in the ENACS clusters to revisit the
question of global vs. local driving of segregation. Since our data
are mostly limited to the central regions of rich clusters (they do
not extend much beyond the virialization radius) our analysis is
largely complementary to those of Goto et al. (2003) and Treu et al.
(2003) whose data go out to much larger projected distances. The paper
is organised as follows. In \S~2 we summarize the data that we used,
in \S~3 we study the morphology-radius relation, and in \S~4 we
investigate the morphology-density relation. Finally, we discuss the
results and summarize our conclusions in \S~5.

\section{Data sample}

\begin{table}
\caption[sample]{The 23 ENACS clusters with galaxy samples \\ 
complete to $M_R$ = -19.5 ($H_0 = 100$ kms$^{-1}$Mpc$^{-1}$). \\
\begin{tabular}{|r|crl|rr|}
\hline
ACO & ${\rm cz_{3K}}$ & ${\rm \sigma_V}$ & Center & ${\rm N_{memb}}$ &
    ${\rm N_{type}}$ \\
\hline
   87   &   16149  &  875 & Geometric     &  17  &   17 \\
  119   &   12997  &  744 & X-ray         &  63  &   62 \\
  151   &   12074  &  399 & Density peak  &  15  &   15 \\
  151   &   15679  &  693 & X-ray         &  37  &   37 \\ 
  168   &   13201  &  524 & X-ray         &  50  &   50 \\
  548E  &   12400  &  706 & X-ray         &  43  &   43 \\
  548W  &   12638  &  819 & X-ray         &  53  &   52 \\
  754   &   16754  &  769 & X-ray         &  38  &   38 \\
  957   &   13661  &  691 & X-ray         &  24  &   24 \\
  978   &   16648  &  497 & cD-galaxy     &  51  &   47 \\
 2040   &   13974  &  602 & X-ray         &  31  &   31 \\
 2052   &   10638  &  654 & X-ray         &  25  &   22 \\
 2401   &   16844  &  475 & cD-galaxy     &  23  &   22 \\
 2734   &   18217  &  579 & X-ray         &  38  &   38 \\
 2799   &   18724  &  493 & cD-galaxy     &  34  &   34 \\
 3122   &   19171  &  780 & Density peak  &  61  &   61 \\
 3128   &   17931  &  809 & X-ray         & 145  &  145 \\
 3158   &   17698  &  977 & X-ray         &  87  &   85 \\
 3223   &   17970  &  597 & cD-galaxy     &  53  &   52 \\
 3341   &   11364  &  561 & X-ray         &  25  &   25 \\
 3528   &   16377  & 1040 & X-ray         &  28  &   28 \\
 3651   &   17863  &  662 & cD-galaxy     &  78  &   78 \\
 3667   &   16620  & 1064 & X-ray         &  99  &   99 \\
\hline
\end{tabular}
\vspace*{0.2cm}

The columns give: the ACO number, the average velocity of the cluster
in the CMBR reference frame (${\rm cz_{3K}}$ in km/s), the global
velocity dispersion of the cluster (${\rm \sigma_V}$ in km/s), the way
in which the center of the cluster was determined, the number of
member galaxies in this sample (${\rm N_{memb}}$), and the number of
member galaxies in this sample with a galaxy type (${\rm N_{type}}$),
either from CCD-imaging and/or from the spectrum.}
\label{sample}
\end{table}

The present discussion is based on data from the ESO Nearby Abell
Cluster Survey (ENACS for short; see Katgert et al. 1996, 1998 -
papers I and V). In order to have a cluster sample that is essentially
volume-limited, we imposed a redshift limit of $z < 0.1$ (see
e.g. paper II, Mazure et al. 1996). Interlopers (non-members) were
eliminated with the interloper removal procedure devised by den
Hartog \& Katgert (1996) as slightly modified by Katgert et al. (2004,
paper XII). We accepted only clusters with at least 20 member
galaxies.

Like Dressler (1980) and Whitmore et al.\ (1993) we applied a limit in
absolute magnitude, which was defined as follows. In paper V the ENACS
spectroscopy was estimated to become significantly incomplete below
$R\sim17$. Using this limit, we find that 33 of our clusters could be
completely sampled down to $M_R = - 19.5$ ($H_0 = 100$
kms$^{-1}$Mpc$^{-1}$). For 21 of these clusters, Katgert et al. (1998)
compared the magnitude distributions of the galaxies with ENACS
redshifts with that of the general galaxy population in the direction
of the clusters, as derived from the EDSGC catalogue produced with
Cosmos (e.g. Collins et al. 1989). From the Cosmos data, 4 of these 21
clusters appeared not to be sampled down to $M_R = - 19.5$, so we
excluded those. The projected galaxy density, in the ENACS dataset, of
the 4 rejected clusters was subsequently used as a guide to the
identification of those 3 clusters among the 12 without COSMOS data,
that are likely to be incomplete down to $M_R = - 19.5$, and which
were therefore excluded.

We used the galaxy types derived in paper VIII, from CCD-imaging
and/or from the ENACS spectrum. Among the selected clusters, there are
3 with galaxy types for less than 80\% of the galaxies, and these were
not used. We are thus left with a sample of 23 clusters, with 1118
member galaxies, for 1105 of which a galaxy type was estimated, and
this cluster sample is described in Table 1.

In paper VIII a full description is given of the classification
method, and we refer to that paper for details. In summary, we used
CCD images of 2295 ENACS galaxies to estimate their morphological
type. In addition, we used the spectral types determined by de Theije
\& Katgert (1999, paper VI) from a PCA/ANN analysis of the ENACS
spectra, after those had been recalibrated with the (mostly new)
morphological types. Finally, we combined all this information
(including also morphological types from the literature), using a set
of calibrated prescriptions for those galaxies with both a
morphological and a spectral type. The inclusion of spectral types is,
strictly speaking, at odds with the terms morphology-radius and
morphology density relation, but as we argued in paper VIII the galaxy
types derived there form a consistent set.

In the present analysis we mostly use the combined morphological and
spectral types, which can be one of the following: E(lliptical), S0
(galaxy), Se (early spiral, i.e. either Sa, Sab or Sb) and Sl (late
spiral, i.e either Sbc, Sc, S/I or I). However, in a few cases, we
will limit ourselves to galaxies with morphological types. Note that
we do not use the galaxies with mixed types (E/S0, S0/S) nor the
generic spirals (Sg). Because it was found, in paper XI, that the
brightest ellipticals (or Eb, i.e. those with $M_R < - 22$) show
luminosity segregation, we did not include those in the present
analysis either, except to confirm their strong radial segregation.
For the sake of consistency we also excluded all galaxies of other
types with \mbox{$M_R < -22$}).
 
In Tab.~\ref{t-stat} we show the number of galaxies within each class,
as well as the number for which the type is morphological. The
spectral types do not add much information for ellipticals and early
spirals, because the spectra of these galaxy types are not very, or
not at all, discriminative. On the contrary, the spectra of S0
galaxies and, in particular, Sl galaxies provide fairly good to very
good discrimination.

\begin{table}
\caption{The number of galaxies with morphological and 
spectral types}
\begin{tabular}{|l|r|r|r|}
\hline
Galaxy type & All & Morph.\\
\hline
Eb &  24 &  24 \\
E  & 149 & 149 \\
S0 & 438 & 337 \\
Se & 130 & 110 \\
Sl & 117 &  55 \\
\hline
\end{tabular}
\label{t-stat}
\end{table}

\begin{figure}[!b]
\includegraphics[width=8.8cm]{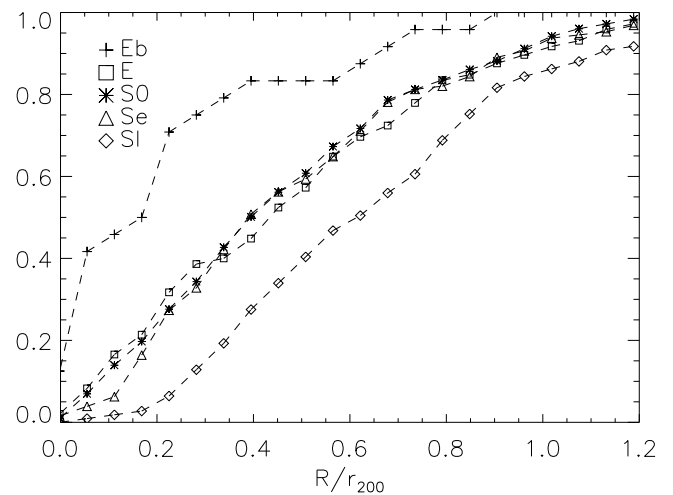}
\caption{The morphology-radius relation. We show the cumulative 
radial distribution for the 5 galaxy types: Eb (ellipticals with $M_R
< -22$), E (other ellipticals), S0 galaxies and early and late spirals
(Se and Sl). The radial distributions of Eb and Sl galaxies are
significantly different from the three other distributions.}
\label{f-MR}
\end{figure}

\section{The morphology-radius relation}

As was already mentioned in \S~1, morphological segregation has two
aspects, viz. one related to global factors and another related to
local conditions. We first analyze the evidence for a global
morphology-radius relation by comparing radial distributions of the
various galaxy classes. We quantify these comparisons through
Kolmogorov-Smirnov (KS-) tests. The KS-test gives the probability,
$P_{KS}$, that two distributions are drawn from the same parent
distribution.

We adopted the center of each cluster as in paper XI. The cluster
center position is either the X-ray center, the position of the
central cD galaxy, the position of the peak in the projected density,
or the geometric center (see Table 1). The projected distance to the
cluster-center, $R$, can be scaled in different ways. Whitmore et
al. (1993, hereafter WGJ) adopted a scale radius within which the
average projected density drops below a certain value.  Instead, we
scaled the cluster-centric radius with $r_{200}$, which is the radius
within which the average density is 200 times as large as the critical
density of the Universe, and which is closely related to the
virialization radius (Navarro et al.\, 1996). Although $r_{200}$
cannot be measured directly from the data, a good approximation is
$r_{200} = \sqrt(3) \sigma_V/(10H(z))$, where $\sigma_{\rm V}$ is the
global velocity dispersion of the cluster and $H(z)$ is the Hubble
parameter at redshift $z$ (see e.g.\ Carlberg et al.\ 1997). The
global velocity dispersion $\sigma_V$ was taken from paper XI and is
listed in Table 1.

In Fig.~\ref{f-MR} we show the morphology-radius relation. Note that
the results in Fig.~\ref{f-MR} use galaxies with morphological {\em
and} spectral types. Dressler et al.  (1980) and WGJ used the E, S0
and S classes in their segregation studies, without subdividing the
ellipticals and spirals, as we do. However, Fig.~\ref{f-MR} clearly
shows that the morphology-radius relation is primarily due to the {\em
brightest} ellipticals (Eb), which are centrally concentrated, and the
{\em late} spirals (Sl) which are almost absent from the central
region ($R > 0.2r_{200}$). There is no evidence that the 'normal'
ellipticals (E), S0 galaxies and early spirals (Se) have different
radial distributions, although their radial distributions are
significantly different from those of bright ellipticals and late
spirals.

\begin{figure}[!b]
\includegraphics[width=8.8cm]{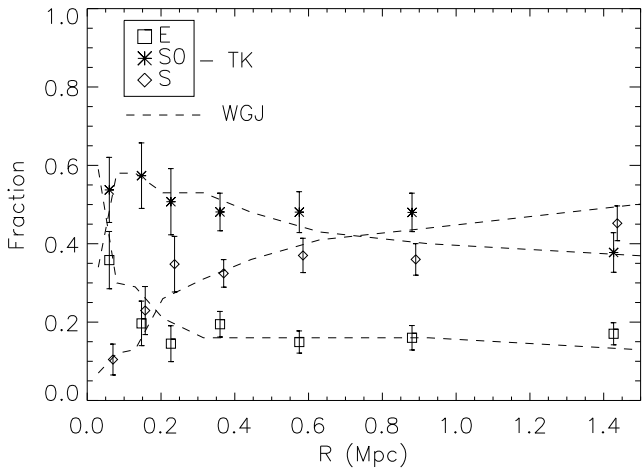}
\caption[fig02]{The morphology-radius relation, expressed as the 
fraction of galaxies of different types for various projected
distances (with $H_0 = 100$ kms$^{-1}$Mpc$^{-1}$) . The dashed lines
were taken from Fig.~4 in WGJ, and the symbols represent our results.}
\label{f-MR-comp}
\end{figure}

In Tab.~\ref{t-MR-KS} we show the results of the KS comparisons
between the various galaxy classes (for all galaxies as well as for
those with morphological types only). If we limit the comparisons to
galaxies with morphological types, we obtain essentially identical
results as when we use all galaxies. Only the Se -- Sl comparison now
yields a KS-probability of 0.03 instead of $ < 0.01 $, probably mostly
due to the much smaller number of Sl involved (cf. Tab.~\ref{t-stat}).

\begin{table}
\caption{The results of the MR comparisons}
\begin{tabular}{|l|r|r|}
\hline
Galaxy samples & \multicolumn{2}{c|}{$P_{KS}$} \\ 
               & All types  & Morph. types \\
\hline
Eb -- E  & $ < 0.01 $ & $ < 0.01 $ \\
Eb -- S0 & $ < 0.01 $ & $ < 0.01 $ \\
Eb -- Se & $ < 0.01 $ & $ < 0.01 $ \\
Eb -- Sl & $ < 0.01 $ & $ < 0.01 $ \\
E  -- S0 & 0.49       & 0.50       \\
E  -- Se & 0.32       & 0.23       \\
E  -- Sl & $ < 0.01 $ & $ < 0.01 $ \\
S0 -- Se & 0.33       & 0.33       \\
S0 -- Sl & $ < 0.01 $ & $ < 0.01 $ \\
Se -- Sl & $ < 0.01 $ & 0.03       \\
\hline
\end{tabular}
\label{t-MR-KS}
\end{table}

In Fig.~\ref{f-MR-comp} we compare our morphology-radius relation with
the one derived by WGJ. Note that in this comparison, we use the
result of WGJ as expressed in $Mpc$ (but corrected to the value of the
Hubble constant that we use), and using our unscaled projected radii
$R$ in $Mpc$. For this comparison we included the brightest
ellipticals in the E class and we combined all spirals, i.e. Se, Sl
and generic spirals. As a result, we have 173 ellipticals, 438 S0
galaxies and 316 spirals (because here we could also include the {\em
generic} spirals). Fig.~\ref{f-MR-comp} shows that the agreement
between the MR-relations of WGJ and ours is quite good, although WGJ
have a slightly higher fraction of spirals in the outer regions, but
not significantly so. Yet, the fact that WGJ had a fainter magnitude
limit (by 0.5 magnitudes), so that they were able to detect relatively
more (faint) Sl galaxies, may be (partly) responsible for the slight
difference. It is especially noteworthy that the agreement for the
ellipticals is also quite good in the central regions (say, for $R <
0.4\ Mpc$). This shows that the segregation of the early-type galaxies
is indeed primarily due to the {\em} brightest ellipticals (which are
largely responsible for the upturn within $\sim 0.1\ Mpc$), even
though there is on average only one of those in each cluster.

\section{The morphology-density relation}

\begin{figure}[!b]
\includegraphics[width=8.8cm]{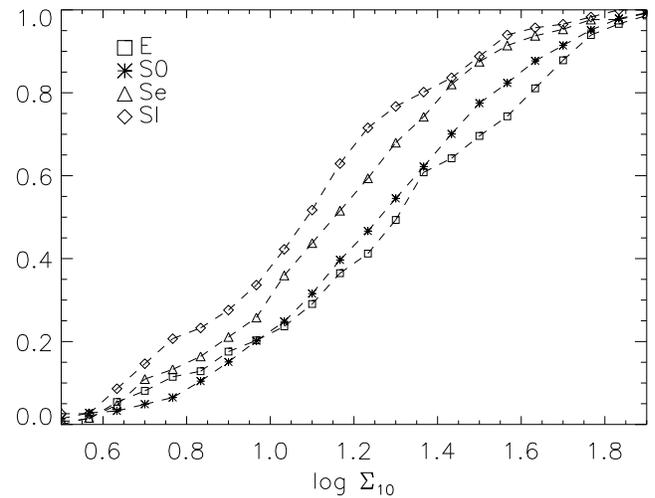}
\caption{The morphology-density relation using $\Sigma_{10}$, the 
projected density derived from the 10-th nearest neighbour. We show
the cumulative distribution of $\Sigma_{10}$ for the various galaxy
classes. Note that the brightest ellipticals are not included in the
elliptical class.}
\label{f-MD-10}
\end{figure}

We now turn to the analysis of the local factors in morphological
segregation, by studying the morphological composition as a function
of projected density. For the determination of the morphology-density
relation, we first followed Dressler's (1980) prescription, i.e., we
used the 10 nearest neighbours (in projection) of each galaxy to
determine the projected density, $\Sigma_{10}$. In Fig.~\ref{f-MD-10}
we show the morphology-density relation, viz. the cumulative
distributions of the galaxies of various types with $\Sigma_{10}$. As
explained before, we did not include the brightest ellipticals. Note
that the results in Fig.~\ref{f-MD-10} use galaxies with morphological
{\em and} spectral types. The results of the KS comparisons of the
$\Sigma_{10}$-distributions are given in Tab.~\ref{t-MR-10-KS}.

Fig.~\ref{f-MD-10} and Tab.~\ref{t-MR-10-KS} show that the
ellipticals, S0 galaxies and early spirals, which have
indistinguishable radial distributions, do not all have the same
distribution of projected density $\Sigma_{10}$. From
Fig.~\ref{f-MD-10} it appears that the average $\Sigma_{10}$ decreases
monotonically from early-type to late-type galaxies. However, only the
comparisons between early and late types (i.e. E or S0 on the one hand
and Se or Sl on the other hand) show a significant difference. If we
limit the comparison to galaxies with morphological types, the Sl
become slighly less different, probably as a result of the smaller
number of Sl involved (see the last column in Tab.~\ref{t-stat}).
However, the general result does not change: early- and late-type
galaxies appear to have different $\Sigma_{10}$-distributions.

\begin{table}
\caption{The results of the $\Sigma_{10}$ MD comparison}
\begin{tabular}{|l|r|r|}
\hline
Galaxy samples & \multicolumn{2}{c|}{$P_{KS}$} \\ 
               & All types  & Morph. types \\
\hline
E  -- S0 & 0.33       & 0.57       \\ 
E  -- Se & $ < 0.01 $ & $ < 0.01 $ \\ 
E  -- Sl & $ < 0.01 $ & 0.02       \\
S0 -- Se & $ < 0.01 $ & $ < 0.01 $ \\
S0 -- Sl & $ < 0.01 $ & 0.09       \\ 
Se -- Sl & 0.07       & 0.96       \\
\hline
\end{tabular}
\label{t-MR-10-KS}
\end{table}

\begin{figure}[!b]
\includegraphics[width=8.8cm]{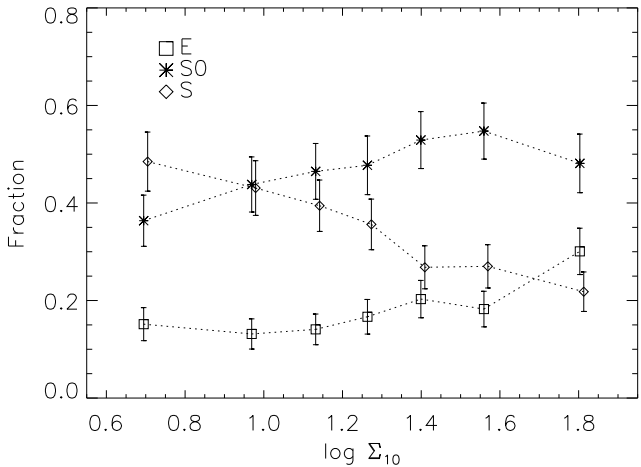}
\caption{The morphology-density relation in the traditional 
representation, i.e. as the variation of the fraction of galaxies of
different morphological types with projected density $\Sigma_{10}$.}
\label{f-MD-dressler}
\end{figure}

In Fig.~\ref{f-MD-dressler} we show our result in the more
traditional fashion, i.e., as the dependence of the fractions of
galaxies of different morphological types on projected density
$\Sigma_{10}$. A detailed comparison of this figure with similar
figures in the literature requires a detailed calibration of the
zero-point of the projected densities. The latter depends on the lower
limit in absolute magnitude, and on the photometric band in which this
is defined. We refrain from a calculation of such zero-point offsets,
but we note that the agreement between our result and that of Dressler
(1980) is very good if our densities were about $10^{0.15}$ smaller
than Dressler's, which is quite plausible.

Comparison with the MDR obtained by Goto et al. (2003), obtained from
the SDSS is even more interesting but, at the same time, less
straightforward. More interesting because Goto et al. also distinguish
early and late disc galaxies, like we do. However, less
straightforward because their morphological types, which were derived
in an automated fashion from the SDSS images, are: early,
intermediate, early disc and late disc. It is not at all trivial to
relate these types to ours, viz. elliptical, S0, early and late
spiral. Judging from their figure 12, and comparing with
Fig.~\ref{f-MD-dressler}, their early-type galaxies could correspond
mostly to our ellipticals.  However, their intermediate-type galaxies
probably represent only a fraction (of the order of two-thirds) of our
S0 galaxies, which leaves the correspondence between their early and
late discs with our early- and late-type spirals ill-defined. 

\begin{figure}[!b]
\includegraphics[width=8.5cm]{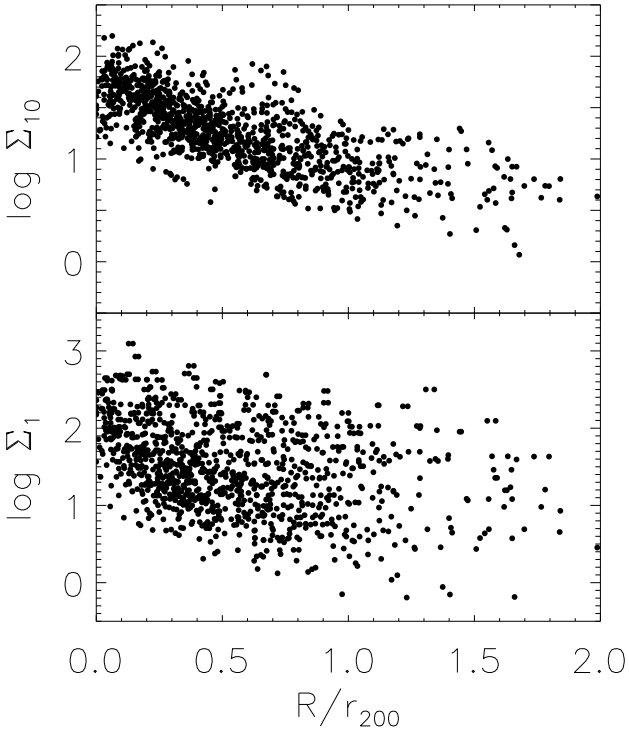}
\caption{The distribution of the galaxies w.r.t $\Sigma_{10}$ 
and $R/r_{200}$ (top) and $\Sigma_{1}$ and $R/r_{200}$ (bottom).
$\Sigma_{10}$ is the projected density derived from the 10-th nearest
neighbour while $\Sigma_{1}$ is derived from the projected distance to
the nearest neighbour.}
\label{f-10-1-R}
\end{figure}

Returning now to Fig.~\ref{f-MD-10}, we stress that it must be
realized that $\Sigma_{10}$ is correlated with radius through the
projected number density profile of the galaxy population. Thus,
$\Sigma_{10}$-distributions of two galaxy samples can be different as
a result of differences in radial distribution. In the upper panel of
Fig.~\ref{f-10-1-R} we show the correlation between $\Sigma_{10}$ and
$R/r_{200}$, which appears to be quite strong.  Apparently,
$\Sigma_{10}$, which was designed to measure the local projected
density, is still a rather global parameter.  Therefore, we defined an
alternative measure of the local projected density as {\bf
$\Sigma_{1}=1/(\pi d^2)$}, where $d$ is the projected distance to the
nearest neighbour. While $\Sigma_{1}$ is more affected by Poisson
noise than $\Sigma_{10}$, the lower panel of Fig.~\ref{f-10-1-R} shows
that it also varies less with $R/r_{200}$ than $\Sigma_{10}$, at least
for $R \ga 0.2 r_{200}$.

One might wonder to what extent $\Sigma_{1}$ might be susceptible to
imperfections in the interloper removal (see Katgert, Biviano \&
Mazure 2004). It is difficult to quantify that in an exact manner, but
from Figure 7 in that same paper, we conclude that the errors in the
interloper removal must be very minor. In addition, the interloper
removal is done without information on galaxy type, so we would expect
these very minor errors to produce random noise in the
morphology-density relation. Below we will discuss the consequences of
the noisy nature of $\Sigma_{1}$.

\begin{figure}
\includegraphics[width=8.8cm]{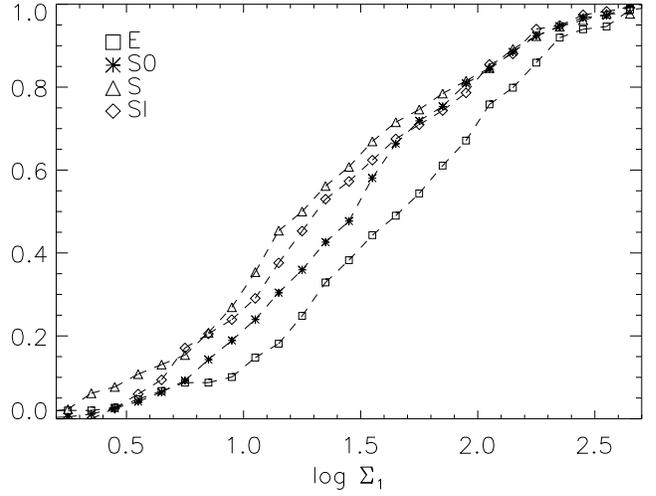}
\caption{The morphology-density relation using $\Sigma_{1}$, the 
projected density derived from the nearest neighbour. We show the
cumulative distribution of $\Sigma_{1}$ for the various galaxy
classes.}
\label{f-MD-1}
\end{figure}

In Fig.~\ref{f-MD-1} we show the morphology-density relation using
$\Sigma_{1}$ instead of $\Sigma_{10}$. Note that we used galaxies with
morphological {\em and} spectral types. As for $\Sigma_{10}$, the
average $\Sigma_{1}$ appears to decrease monotonically from early-type
to late-type galaxies, except for the late spirals, which are
intermediate between the S0 galaxies and the early spirals. The
results of the KS comparisons are given in Tab.~\ref{t-MR-1-KS}. The
late spirals are indeed not very different, if at all, from the other
three classes, and certainly {\em much less} so than in the
morphology-radius relation. The other three galaxy classes are found
to have significantly different $\Sigma_{1}$-distributions. It is
important to realize that this result cannot be affected by cross-talk
from the morphology-radius relation, since the radial distributions of
E, S0 and Se galaxies were found to be indistinguishable.

In view of the novelty of $\Sigma_{1}$, and the rather large
noise in it, we have checked the robustness of our conclusions. We
have done this by repeating our analysis for a set of 1000 azimuthal
scramblings of our cluster sample. By leaving the radial distribution
unchanged, we have avoided introducing unwanted cross-talk from the
MR-relation. At the same time, the azimuthal scrambling will destroy
the relations between morphological type and local projected density,
as found in Fig.~\ref{f-MD-1} and Tab.~\ref{t-MR-1-KS}. In other
words: if the strong dissimilarities of the $\Sigma_{1}$-
distributions of ellipticals, S0 galaxies and early spirals are real
we would expect that in the scrambled data the low values of the
KS-probabilities that we observed ($P_{KS} < 0.01$) are very rare.

The results of the 1000 scramblings indeed fully confirm this
expectation. Only in 2 out of 1000 cases does the E -- S0 comparison
give $P_{KS} < 0.01$, while for the E -- Se and the S0 -- Se
comparisons the corresponding fractions are 28 and 29 out of
1000. This result indicates that notwithstanding the fairly large
Poisson noise in $\Sigma_{1}$, our results about segregation in
$\Sigma_{1}$ are robust.

\begin{table}
\caption{The results of the $\Sigma_{1}$ MD comparison}
\begin{tabular}{|l|r|r|}
\hline
Galaxy samples & \multicolumn{2}{c|}{$P_{KS}$} \\ & All types &
               Morph. types \\
\hline
E  -- S0 & $ < 0.01 $ & $ < 0.01 $ \\ 
E  -- Se & $ < 0.01 $ & $ < 0.01 $ \\
E  -- Sl & $ < 0.01 $ &    0.12    \\ 
S0 -- Se & $ < 0.01 $ & $ < 0.01 $ \\
S0 -- Sl &    0.19    &    0.96    \\
Se -- Sl &    0.50    &    0.03    \\
\hline
\end{tabular}
\label{t-MR-1-KS}
\end{table}

From the cumulative distributions shown in Figs.~\ref{f-MR},
\ref{f-MD-10} and \ref{f-MD-1} we conclude that the various classes of
galaxies obey different segregation rules. It is evident that
position in the cluster (i.e projected distance from the center) is
the main factor that sets the brightest ellipticals and the late
spirals apart. On the contrary, the differences between ellipticals,
S0 galaxies and early spirals are most apparent in their distributions
of projected density, either $\Sigma_1$ or $\Sigma_{10}$, or both. The
segregation of ellipticals, S0 galaxies and early spirals is therefore
probably driven primarily by local conditions, while that of late
spirals and brightest ellipticals seems primarily driven by global
conditions.

\section{Discussion and conclusions}

For about 850 galaxies in 23 ENACS clusters we studied morphological
segregation in projected radius and projected density. The sample of
galaxies is complete to a magnitude $M_R=-19.5$ ($H_0 = 100$
kms$^{-1}$Mpc$^{-1}$). 

Our analysis has yielded two main results. First, the distribution of
projected radius (i.e. the morphology-radius relation) shows that the
brightest ellipticals (i.e. those with $M_R < -22$) and the late
spirals have distributions that are significantly different from those
of the other ellipticals (with $M_R \ge -22$), the S0 galaxies and the
early spirals. The latter three galaxy classes have indistinguishable
radial distributions, which are intermediate to that of the brightest
ellipticals (very centrally concentrated, with ~ 75\% of the brightest
ellipticals within $0.3\ r_{200}$) and that of the late spirals (of
which only ~ 15\% have $R < 0.3\ r_{200}$).

Secondly, the morphology-density relation shows that the ellipticals,
S0 galaxies and early spirals have significantly different
distributions of local density $\Sigma_{1}$. On average, ellipticals
prefer environments where the density is highest, while early spirals
avoid these environments. The behaviour of S0 galaxies is
intermediate; they are present in low density as well as high-density
environments. The fact that the ellipticals and S0 galaxies have
indistinguishable distributions of the less local density
$\Sigma_{10}$, is due to the significant correlation between
$\Sigma_{10}$ and projected radius $R$.

The first result suggests that for the brightest ellipticals and for
the late spirals global effects, such as position in the cluster, are
more important than the properties of the local environment. On the
contrary, the second result suggests that for the ellipticals, S0
galaxies and early spirals the position in the cluster is much less
important than the local conditions. Note that the latter result does
not suffer from cross-talk from the radial distribution, as the three
classes have essentially identical $R$-distributions.

Although radial segregation was observed before, the new result of our
analysis is that only {\em the brightest} ellipticals and the {\em
late} spirals show segregation, while the other ellipticals and the
early spirals do not. In the SDSS data discussed by Goto et al. (2003)
a similar difference between what they refer to as `early and late
disks' is visible. On the other hand, the strong increase towards the
center of the fraction of ellipticals found by Goto et al. is probably
mostly due to the fact that they do not consider the brightest
ellipticals separately, as should be done (see paper XI).

The present analysis shows that the distinction between global and
local segregation is not simply a matter of inner regions vs. outer
ones, as one might have concluded from the results obtained by
Dominguez et al. (2001). The different segregation `rules' that they
find for the inner and outer regions appear to be manifestations of
different segregation behaviour of the various types of galaxies.

Our conclusions provide confirmation of several current ideas about
galaxy evolution and transformation in clusters of galaxies. These
ideas distinguish between two different kinds of processes: formation
of galaxies through mergers of smaller galaxies, or transformation of
galaxies through encounters with other galaxies or by the influence of
the cluster potential. We now describe briefly how our results may
give information about these processes, taking the several galaxy
classes one at a time, from early to late Hubble types.

The segregation of the {\em brightest ellipticals} was investigated by
several authors (e.g.\ Rood \& Tunrose 1968, Capelato et al.\ 1981,
Kashikawa et al.\ 1998, Beisbart \& Kesher 2000 and in paper XI). Those
studies indicate that the brightest ellipticals have been (and are
being) formed probably by merging and accretion in the central regions
of clusters (see e.g.\ Governato et al.\ 2001). Global estimates of
the time-scale involved in the accretion, viz. that of dynamical
friction, show that only in the central regions this time-scale is
sufficiently short that this process may be important.

Most of {\em the other ellipticals} have probably formed by merging of
disk galaxies (e.g.\ Toomre \& Toomre 1972, Barnes \& Hernquist 1996,
Aguerri et al.\ 2001). Direct evidence for mergers was found in
high-redshift clusters (e.g.\ Lavery \& Henry 1988; Lavery et
al. 1992; Dressler et al.\ 1994, Couch et al.\ 1998; van Dokkum et
al. 1999). In the hierarchical scenario, the formation of ellipticals
thus takes place in relatively dense regions (proto-clusters) where
there were enough objects that could merge. Therefore, it is not
surprising that we find few ellipticals in regions with low projected
densities.

The {\em S0 galaxies} and {\em early spirals} must be discussed
together as they are likely to be related through transformation
processes. Several mechanisms are thought to be important in the
evolution and transformation, such as the stripping of gas, impulsive
tidal interactions between galaxies and mergers. These have been
described in papers by e.g. Moore et al. (1998, 1999), Abadi et al.
(1999) and Okamoto, \& Nagashima (2001). It appears that impulsive
encounters of early spirals with other galaxies may lead to stripping
of a modest fraction of the stellar component and an increase of the
vertical scale-height of the disk.

Several studies have shown that S0 galaxies, like ellipticals, are
passively evolving galaxies, which mainly contain stars formed at high
redshifts (e.g.\ Bower et al.\ 1992; Ellis et al.\ 1997; Lucey et
al. 1991, van Dokkum et al.\ 1996, 1998). However, it should be
remembered that shorter luminosity- weighted ages were found for faint
S0 galaxies (e.g.\ Smail et al. 2001). At the same time, evidence has
accumulated that the fraction of S0 galaxies in clusters has increased
strongly since $z=0.5$ (Dressler et al.\ 1997; Fasano et al.\ 2000),
and this is generally thought to be due to a transformation from early
spirals into S0 galaxies.

Poggianti et al.\ (1999) discuss the evidence for spectral and
morphological transformations of early spirals into S0 galaxies. At
intermediate redshifts starformation in spirals is probably quenched
after a final starburst (e.g.\ Dressler \& Gunn 1983; Couch \&
Sharples 1987), which leads to a spectral transformation. This process
occurred when galaxies fell into the cluster (Dressler et al. 1999;
Poggianti et al.\ 1999; Ellingson et al.\ 2001). The process that
transformed early spirals into S0 galaxies probably occurred later and
on longer time-scales (see also Poggianti et al.\ 1999; Jones et
al. 2000). One process by which the starformation could be quenched is
the removal of the gas in spiral galaxies by ram pressure and
turbulent or viscous stripping through the hot intra-cluster medium
(Quilis et al. 2000).

Harassment and impulsive encounters (Moore et al.\ 1998) are most
likely the processes by which early spirals can be transformed into S0
galaxies. This is supported by our finding that, on average, the local
density around early spirals is somewhat smaller than that around S0
galaxies. The transformation efficiency is likely to be larger if the
density is higher, and this would lead to a selection against early
spirals in higher density environments. Biviano \& Katgert (2004,
paper XIII) studied the velocity distributions of the various galaxy
classes and concluded that those also provide marginal support for
this picture.

Finally, while the brightest ellipticals are found exclusively in the
central regions of clusters, the {\em late spirals} avoid those
regions almost completely. This suggests that the late spirals are
probably destroyed by the tidal forces of the cluster potential. As
shown by Moore et al. (1999), the fate of spiral galaxies in the
central regions of clusters depends very much on the 'hardness' of
their gravitational potential. The `destruction hypothesis' for late
spirals is therefore very plausible because their rotation curves
indicate that their mass distributions are much less centrally
concentrated that those of early spirals (e.g. Corradi \& Capaccioli
1990, Biviano et al. 1991, Adami et al. 1999 and Dale et al. 2001).

We refrain from estimating relevant timescales and efficiencies of the
various processes mentioned here. However, we note that Treu et
al. (2003) have made such estimates by defining three distinct regimes
in a cluster according to the different physical processes that drive
the various types of segregation.

\begin{acknowledgements}
We thank Andrea Biviano for a careful reading of the manuscript and
for useful suggestions.
\end{acknowledgements}

\vfill
\end{document}